# Optical power meter using radiation pressure measurement

Patrick PINOT[*] and Zaccaria SILVESTRI

Laboratoire commun de métrologie LNE/CNAM 61 rue du Landy F93210 La Plaine-Saint-Denis, France

Abstract

This paper describes a radiation pressure meter based on a diamagnetic spring. We take advantage of the diamagnetic property of pyrolytic carbon to make an elementary levitated system. It is equivalent to a torsional spring-mass-damper system consisting of a small pyrolytic carbon disc levitated above a permanent magnet array. There are several possible measurement modes. In this paper, only the angular response to an optical power single-step is described. An optical detection composed of a laser diode, a mirror and a position sensitive detector (PSD) allow measurement of the angular deflection proportional to the voltage delivered by the PSD. Once the parameters of the levitated system depending on its geometrical and physical characteristics have been determined regardless of any optical power, by applying a simple physical law, one can deduce the value of the optical power to be measured from the measurement of the first maximum of the output voltage amplitude.

## 1. Introduction

High power lasers (of a few kilowatts) are used in a wide range of industrial manufacturing processes ranging from welding, cutting, marking and additive manufacturing of structures made of various materials including metals. The laser is a working tool with a long lifetime, can be included in monitoring and control systems based on intelligent sensing techniques, and allows zero-fault production. This makes it necessary to improve high optical power measurement. At the same time, intermediate optical powers (from a few milliwatts to a few watts) are used in different fields such as telecommunications, alignment, laser shows, microlithography and energy sources.

The two most common techniques used to measure optical power use either non-thermal sensors based on photo-electron interaction (photodiodes for example) or thermal sensors based on thermal energy absorption (thermopile sensors). Photodiodes can be used to measure the high power by sampling only a fraction of the optical power only if the small fractional splitting ratio is known. In addition, photodiodes are sensitive to temperature and the optical power distribution. Thermopile sensors can measure high-power directly, but are somewhat slow and require the full laser power be incident on the power meter. These two measurement techniques are exclusive operations.

Measurements of laser power using radiation pressure have been reported recently [1–8]. In particular, Williams *et al.* [8] have developed a portable, high-accuracy, non-absorbing laser power measurement at kilowatt levels.

The work presented hereafter is a preliminary study of a laser power meter based on pyrolytic carbon levitation to detect the moment of force provided by radiation pressure of a laser beam. The optical power range studied is from 100 mW to 1 W. We use the strong diamagnetism of pyrolytic carbon (PyC) to carry out 6D-stabilisation of a PyC disc levitated at ambient temperature above a magnet array.

The rest of this paper is structured as follows. Section 2 briefly presents the theoretical principle of a spring-mass-damper system and an experimental set-up made from very simple and inexpensive elements. Section 3 provides some theoretical considerations used to specify the main characteristics of the system. Preliminary results are presented in Section 4. Section 5 gives some potential sources of error and uncertainties in measurement which are discussed before the conclusion in Section 6.

## 2. Principle of the spring-mass-damper system and experimental set-up

### 2.1. Radiation pressure

The concept of radiation pressure grew in the late nineteenth and early twentieth centuries starting from Maxwell's theory of electromagnetism [9,10]. Later, with the advent of quantum theory, this concept allowed one to define the linear photon momentum $p = \hbar\omega/c$ where $\hbar$ is the reduced Planck constant, $\omega$ the photon frequency, and $c$ the speed of light *in vacuo*. A perfect mirror reflecting a photon at normal incidence reverses its momentum. Consequently, the force $F_R$ exerted by $dn$ photons reflected at normal incidence on a perfect mirror is equal to $dp/dt = 2\hbar\omega dn/cdt$. The equation $\hbar\omega dn/dt$ corresponds to the optical power $P$ of the incident light. The force $F_R$ is given by:

$$F_R = \frac{2P}{c} \tag{1}$$

*Corresponding author: patrick.pinot@cnam.fr

Measuring optical power via a radiation pressure measurement has been suggested previously [11–13].
In a real case, the surface is neither plane nor perfectly reflecting. One part of the incident photons is reflected and the other is absorbed. Let $\alpha$ be the fraction of absorbed photons and $R$ the reflectivity of the mirror. The resultant coefficient $r$ of reflectivity and absorption is $r = R + (1-R)\alpha/2$.
In addition, the incident photon beam is never perfectly normal to the mirror surface. If $\theta$ is the angle of incidence, the force component along the normal to the mirror is proportional to $\cos\theta$. Thus Equation (1) becomes:

$$F_R = \frac{[2R + (1-R)\alpha]P\cos\theta}{c} = \frac{2r\,P\cos\theta}{c} \tag{2}$$

2.2. Magnetic spring

Pyrolytic carbon has interesting thermal, optical and electrical properties that can be used for various applications. In this work, however, we focus on its strong diamagnetism used to maintain a PyC disc levitated in a horizontal $x$-$y$ plane at a height $z_L$ above a magnet array made of cubic NdFeB magnets.
The magnetic interaction force between the PyC disc and the magnet array allows the disc to be levitated and acts as a magnetic spring with six degrees of freedom. However, the force components in the six degrees of freedom are different. They are functions of the PyC volume magnetic susceptibility $\chi$, the imperfections of the PyC disc and the distribution of the magnetic flux density $B$ provided by the magnets. Given that the PyC magnetic susceptibilities $\chi_{x-y}$ and $\chi_z$ are respectively -8.5×10$^{-5}$ and -4.5×10$^{-4}$ for the pyrolytic carbon used here, the restoring magnetic force gradients acting on the PyC disc are very different in the $x$-$y$ plane and along the $z$-direction.
In this work, we use as perfectly circular a PyC disc as possible (about 20 mm in diameter and 1.0 mm thick). The magnet array is made of nine cubic grade N42 NdFeB magnets (edge $a = 10$ mm). The manufacturer specifies a remanent induction $B_r$ for each magnet between 1.29 and 1.32 T. Under these conditions, the PyC disc remains at a stable levitation position in the horizontal plane. The levitation height $z_L$ between the magnet array surface and the lower face of the PyC disc supporting various mechanical and optical elements is about 0.32 mm (measured using an optical method). Consequently, the levitation height at the middle of the PyC disc where the magnetic force acts is $z_0 \approx 0.82$ mm.
Provided the magnetic flux density $B$ is vertical, homogeneous and constant in any horizontal plane $x$-$y$, a homogeneous piece of diamagnetic PyC can slide freely in the $x$-$y$ plane at its levitation height $z$. The levitation force $F_z$ is given roughly by:

$$F_z = -\chi_z \frac{A}{2\mu_0} \int_z B \frac{\partial B}{\partial z} dz \tag{3}$$

where $A$ is the area of the sheet of material (PyC) and $\mu_0$ the vacuum permeability ($\mu_0 = 4\pi \times 10^{-7}$ H m$^{-1}$).
This force balances the weight $mg$ of the mass $m$ of the levitated system subjected to the acceleration due to gravity $g$.
In fact, the magnet array is composed of a juxtaposition of small identical permanent magnets alternating vertically south and north poles in the horizontal $x$-$y$ plane. Consequently, the $B$ field in the $x$-$y$ plane at a height $z$ above the magnet array exhibits a quasi 2D periodicity. As explained by Simon and Geim [14], the stability conditions to levitate an object in a vertical magnetic field $B$ are $\chi < 0$ (volume magnetic susceptibility of the object), $\partial^2 B^2/\partial x^2 > 0$ and $\partial^2 B^2/\partial y^2 > 0$ (horizontal stability) and $\partial^2 B^2/\partial z^2 > 0$ (vertical stability). Consequently, the PyC disc remains at one of its stable positions in the horizontal plane. In his PhD thesis, Barrot [15] calculates the magnetic field created by different arrangements of permanent magnets and shows images obtained by a simulation software for the 3D distribution of the $B$-field above a magnet array composed of nine cubic NdFeB magnets in different orientations. While this gives an idea of the magnetic field distribution for our configuration, the magnet size and the arrangement differ slightly from those of our device.

2.3. Apparatus

Among the different possible configurations for developing an optical power meter based on a diamagnetic spring, we have chosen to develop the rotational one, designated hereafter by the letters LPM (Laser Power Meter).

Figure 1 shows a schematic diagram of the experimental set-up made up of four main parts.

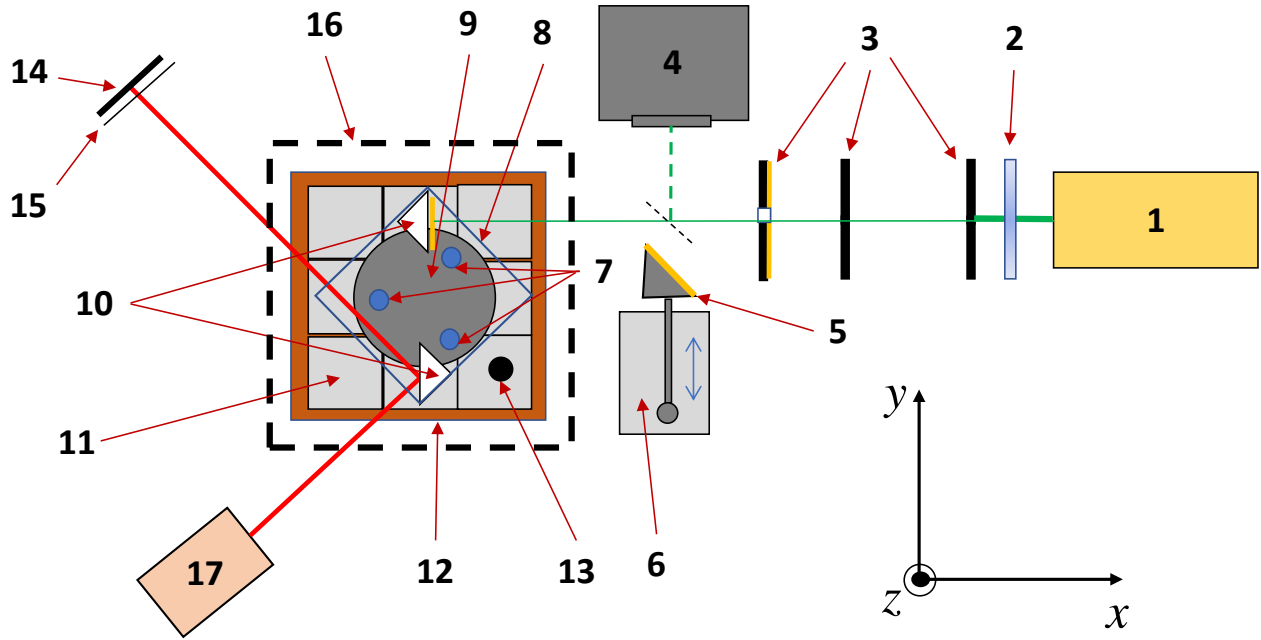

Figure 1: Schematic diagram of the experimental set-up.
 1 - Laser under test ($\lambda$ = 1064 nm); 2 – High-speed shutter; 3 – Diaphragms (Ø = 1.5 mm); 4 - Integrating sphere photodiode power sensor; 5 - Retractable golden mirror; 6 - Translation stage; 7 – Polyamide balls (Ø = 3 mm; $m$ = 15 mg); 8 – Cover glass (18 mm × 18 mm × 0.17 mm; $m$ = 112 mg); 9 - PyC disk (Ø = 20 mm; $t$ = 1 mm; $m$ = 657 mg); 10 – Right angle prism mirrors (coated for the wavelength; $m$ = 158 mg); 11 – Nine-NdFeB-magnet array (10 mm ×10 mm × 10 mm); 12 - Peltier module; 13 – 10 k$\Omega$ thermistor; 14 - PSD sensor; 15 - Interference filter ($\lambda$ = (633 ± 10) nm); 16 – Box to protect against radiation and draughts; 17 - Laser diode ($\lambda$ = 635 nm).

The first consists of two diaphragms with an aperture of about 1.5 mm. A gold coated third diaphragm with an aperture of 3 mm protects the device against the radiation emitted by any overheating of the two first diaphragms (here the laser under test has a wavelength $\lambda$ = 1064 nm and a maximum optical power $P_{max} \approx 1$ W). A high-speed shutter is placed between the laser head and the first diaphragm and used with its controller (Thorlabs SHB1) to control the exposure time. The angle of incidence of the laser beam under test is not perfectly normal to the mirror surface so as to avoid any reflection on the path of the incident beam. The second part allows us to measure the power of the laser beam passing through the diaphragms via a 45° retractable mirror which reflects the beam onto an integrating sphere photodiode power sensor (Thorlabs S142C - 350 - 1100 nm, 5 W). The reflection coefficient of the mirror for 45° incidence at 1064 nm is 0.99. The third is the sensitive element consisting of the PyC disc on which three polyamide balls are placed in three small holes forming an equilateral triangle supporting a cover glass. Two right angle prism mirrors are placed symmetrically on the cover glass. One is coated on its largest face to reflect the wavelength of the laser under test (reflectance at 8° 0.98 at 532 nm and 0.991 at 1064 nm) and the other to reflect the wavelength of the laser diode (635 nm). The total mass $m$ of the levitated system is 1.18 g.

The magnetic flux density, generated by the magnet array, depends greatly on the temperature of the magnets. This temperature, measured by an NTC thermistor, is regulated by a Peltier module driven using dedicated software. It is kept within ± 4 mK of ambient temperature. The forth part serves to detect the angular deflection of the levitated system exposed to the laser beam under test. For this, a laser diode beam ($\lambda$ = 635 nm) is reflected by a right-angle prism mirror onto a position sensitive detector (PSD) using a PIN diode. The sensor is protected from other wavelengths by an interference filter for 635 nm.

The sensitive part of our device is housed in a small box with three holes, one for the IR beam of the laser under test, another for the detection laser beams and a third for the IR beam reflected by the right-angle prism mirror. The same box protects the PyC disc against draughts and spurious IR reflections. It also improves the temperature stability of the PyC disc. The sensitivity of the magnetic spring depends not only on the horizontality of the magnet array, the symmetrical magnetic flux density distribution, the radial runout of the PyC disc and the lever arm between the impact point of the laser beam and the centre of rotation of the levitated system, but also on whether the centre of gravity of the levitated system and the centre of rotation of the PyC disc lie on the same vertical axis so as to limit the influence of any unwanted linear and angular motions.

## 3. Theoretical considerations

### 3.1. Magnetic flux density

Equation (4) allows one to calculate the magnetic flux density $B$ at a height $z$ and its gradient $\partial B/\partial z$. It is valid around the $z$-axis for a single cubic magnet (for $\beta = 1$). However, the magnetic field distribution for our set-up composed of nine juxtaposed magnets is quite complex. For a given height, the mean value of the magnetic flux density of the nine-magnet array is a little greater than the value estimated for one magnet. For instance, De Pasquale *et al*. [16] have shown the magnetic field distribution of a four-magnet array calculated by a finite element modelling (FEM). They explained that, given the diamagnetic permeability of the graphite, the omission of the levitated PyC system in the simulation does not introduce appreciable errors in the estimation of the magnetic field distribution. As an illustration of the 2D periodicity of $B_z$, Figure 2 shows a simulation made for our magnet configuration for $z = 0$ using the ANSYS AIM 18.1 academic simulation software.

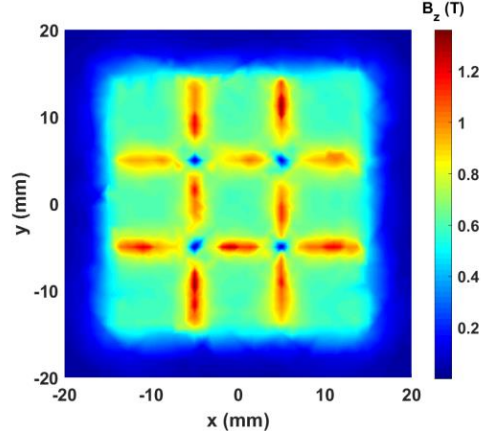

Figure 2: Simulation of $B_z$ for the nine-magnet array at $z = 0$ made using the ANSYS AIM academic software (version 18.2).

Using equation (4) to calculate the mean value of $Bz$ at the height $z$, we introduce a weighting factor $\beta$ in the following equation:

$$B_z = \beta \frac{B_r}{\pi} \left[ \arctan\left( \frac{a^2}{2z_i \sqrt{4z_i^2 + 2a^2}} \right) - \arctan\left( \frac{a^2}{2(a+z_i)\sqrt{4(a+z_i)^2 + 2a^2}} \right) \right] \quad (4)$$

where $B_r = 1.3$ T, $a = 10$ mm and $\beta = 1.6$.
We obtain $B_1 = 0.69$ T for $z_1 = 1.4$ mm and $B_2 = 0.85$ T for $z_2 = z_L = 0.32$ mm.
Note that the value $\beta$ is chosen to obtain a calculated value of $z_0$ as close to the measured one as possible.

### 3.2. Linear spring-mass-damper system

To justify the choice of developing a system based on torsional oscillations around the rotation axis of the PyC disc, we examine other possible configurations.
First, we assume the PyC disc has only a single degree of freedom along the vertical $z$-axis. When the vertical light beam power $P$ is modulated as $P_0 \cos\omega t$, the system is driven by a sinusoidal force $F_R$ caused by the laser beam hitting the mirror and modulated at the same angular frequency $\omega$.
The equation of motion for the forced damped harmonic oscillator is:

$$m \frac{d^2 z}{dt^2} + \gamma_z \frac{dz}{dt} + k_z z = F_0 \cos\omega t \quad (5)$$

where $k_z$ is the spring constant for translation along the $z$-axis, $\gamma_z$ the mechanical resistance (or damping constant) and $F_0$ the amplitude of the modulated force $F_R$. Here, the centre of gravity of the levitated system lies on its symmetry axis, as does the line of incidence of the laser beam.

### 3.2.1. Spring constant $k_z$

In equation (5), $k_z$ corresponds to the magnetic interaction force.
If we assume the gradient $\partial B/\partial z$ is constant for small displacements around the balanced levitation height then, using equation (4), we can estimate the gradient as $\Delta B/\Delta z$ where $\Delta B = B_1 - B_2$ and $\Delta z = z_1 - z_2$ corresponding to the thickness of the PyC disc. Then, using equation (2) and introducing the disc radius $R_d$, we obtain an approximate equation for the restoring magnetic force gradient:

$$k_z = \frac{dF_z}{dz} \approx -\chi_z \frac{\pi R_d^2 B_z \Delta B_z}{\mu_0 \Delta z} \tag{6}$$

This yields $k_z \approx 14$ N m$^{-1}$. The balanced levitation height $z_0$ is deduced from the following equation:

$$z_0 \approx \frac{mg}{k_z} \tag{7}$$

where $g \approx 9.81$ m s$^{-2}$. Note that the centre of gravity of the levitated system lies about 2.8 mm above the magnet array surface, but the magnetic force acts on the PyC disc whose the centre lies a distance $z_0$ above the array.

Equation (7) yields $z_0 = 0.822$ mm, close to its measured value (0.82 mm), the relative measurement uncertainty of which is a few percent. Of course the experimental value of $k_z$ should be determined directly from equation (7) using the measured value of $z_0$, but the method using equations (4) and (6) to calculate $k_z$ and equation (7) to verify the value $z_0$ allows us to obtain information on the approximate values of $B_z$ in the vicinity of the PyC disc.

### 3.2.2. Mechanical resistance $\gamma_z$

Any analysis of sensors having an air film between two movable parts [17] must include Brownian motion. Our device can be modelled as a spring-mass-damper system with Brownian noise due to air damping. The spectral density of the force related to any mechanical resistance $\gamma$ is given from Nyquist's relation even if the resistance is a function of frequency [18].
Gabrielson [16] gives the following equation for the Brownian force equation in newtons per root hertz:

$$F_B = \sqrt{4kT\gamma_z} \tag{8}$$

where $k$ is the Boltzmann constant and $T$ the thermodynamic temperature of air.

Starr [21] has calculated the equivalent mechanical resistance for a thin circular disc of radius $R_d$, with a nominal air film thickness $z_L$:

$$\gamma_z = \frac{3\pi\mu R_d^4}{2z_L^3} \tag{9}$$

where $\mu$ is the dynamic viscosity (18×10$^{-6}$ kg m$^{-1}$s$^{-1}$ for air at 20 °C).

Equation (9) gives $\gamma_z \approx 2.6 \times 10^{-2}$ kg s$^{-1}$. Note that the presence of other mechanical elements (a cover glass for instance) can increase this value.

### 3.2.3. Resonant frequency
#### 3.2.3.1. Translation along $z$-axis

Without friction, the natural angular frequency $\omega_0$ of the undriven undamped oscillator is given by:

$$\omega_0 = \sqrt{\frac{k_z}{m}} \tag{10}$$

Here, we obtain $\omega_0$ = 108.9 rad s$^{-1}$ (*i.e.* a natural frequency of 17.3 Hz). The resonance (maximum value of the amplitude) occurs when:

$$\omega_r = \sqrt{\omega_0^2 - \frac{1}{2}\left(\frac{\gamma_z}{m}\right)^2} \tag{11}$$

The calculation with the numerical values of equation (11) gives $f_r \approx$ 17.2 Hz ($\omega_r \approx$ 107.8 rad s$^{-1}$) with a relative uncertainty of about 10%. This value is consistent with experimental observation.

For the lightly-damped oscillator, $\omega_0 \gg \gamma/2m$, the resonance occurs when $\omega \sim \omega_0$ and the amplitude $z_A$ at resonance is:

$$z_A \approx \frac{F_0}{\gamma_z \omega_0} \approx \frac{[2R+(1-R)\alpha]\cos(\theta)P_0}{c\gamma_z \omega_0} \tag{12}$$

Equation (12) suggests the resonance amplitude is roughly proportional to the optical power.

The ratio $\gamma_z/2m$ is about 11 s$^{-1}$. Consequently, the system can be regarded as lightly-damped which makes the resonance amplitude $z_A$ about 3 nm for $r$ = 1, $\theta$ = 0 and $P_0$ = 1 W. Such a tiny amplitude is too difficult to measure accurately and would be buried in the noise, so this method was abandoned.

3.2.3.2. Translation in plane *x-y*

As a comparative example, the same type of estimate can be made for the horizontal diamagnetic spring. Equation (5) is modified by replacing *z* by *x* (or *y*) with the direction of an external force along the *x*-axis (or *y*-axis). The theoretical calculation requires FEM, without which one must rely on experimental observations of the system's behavior.

We can determine experimentally the logarithmic decrement $\delta$ that characterises the rate at which the oscillations of the system are damped:

$$\delta = \ln\left(\frac{\varphi_{t\max}}{\varphi_{t\max+T}}\right) = \kappa T_d \tag{13}$$

where $T_d$ is the pseudo-period and *t*max the time at which a maximum of angular amplitude occurs.

From three experimental observations, using equation (13) we have obtained an approximate mean value for $\delta$ of 2.90 and a mean value for $\kappa$ of about 2.59 s$^{-1}$ with $T_d \approx$ 1.12 s ($f_d \approx$ 0.89 Hz, *i.e* $\omega_d \approx$ 5.61 rad s$^{-1}$).

The angular frequency of a damped oscillation $\omega_d$ is given by:

$$\omega_d = \sqrt{\omega_0^2 - \kappa^2} \tag{14}$$

For the natural angular frequency of oscillation along the *x*-axis (or *y*-axis), we find $\omega_0 \approx$ 6.18 rad s$^{-1}$. Using equation (10), we estimate $k_x \approx 43\times10^{-3}$ N m$^{-1}$. As expected, the horizontal stiffness is much smaller than the vertical one, but since the mechanical resistance $\gamma_x$ is about 1.8 kg s$^{-1}$, the damping is much larger. The accuracy of this method is limited to a few tens of percent. For instance, a horizontal optical force of 6.7 nN corresponding to an optical power of 1 W should displace the levitated system by 0.16 µm. While we would expect a relative measurement uncertainty of 10% or less, this configuration far too sensitive to seismic vibrations to make it useful in practice.

3.3. Torsional spring-mass-damper system

The aforementioned experimental and theoretical considerations on the use of linear translation of magnetic spring subjected to a radiation pressure show their limits and led to our studying torsional configuration similar to a torsion balance.

3.3.1. Angular spring constant

In equation (5), if we assume the force moves the target mirror horizontally by a distance $l_m$ relative to the centre of gravity of the levitated system (the system is balanced by placing symmetrically another mirror), the radiation pressure on the mirror will generate a moment on the levitated system. If in addition the system undergoes no translation (*i.e.* there is no translation of the centre of rotation in the horizontal plane) and given the disc is imperfect, there will be a non-zero angular spring constant. The equation of motion for the forced damped harmonic oscillator becomes:

$$I\frac{d^2\varphi}{dt^2} + \gamma_\varphi \frac{d\varphi}{dt} + k_\varphi \varphi = F_0 l_m \cos(\varphi)\cos(\omega t) \tag{15}$$

where $I$ is the moment of inertia.

Given we know the mass, size and position of the centre of gravity of each element of the levitated system. We can calculate its moment of inertia for rotation about its centre by adding the moments of inertia of the PyC disc $I_d$, the three polyamide balls $I_b$, the cover glass $I_c$ and the two right-angle prism mirrors $I_m$.
Its standard uncertainty is the quadratic sum of the standard uncertainties of each component.

Moreover, the magnitude of the torque depends on the radiation force $F_R$ applied, the length $l_m = 11$ mm of the lever arm connecting the axis to the point of force application, and the angle between the force vector and the lever arm. If the force vector (incident laser beam) is normal to the mirror surface, the moment of force for a laser power of 1 W is a mere $74 \times 10^{-12}$ N m.
At the beginning, we machined a PyC disc as circular and perfect as possible. This turned out not to be a good idea because the rotational stiffness was too weak and the angular frequency too low to determine easily. Under these conditions, the angular position of the PyC disc wobbled due to draughts, so the deviation due to the irradiation force was indistinguishable from the noise. To fix the ball position, we drilled three holes (Ø= 1.5 mm) in the PyC disc. This was sufficient to provide a steady angular position with a non-negligible angular spring constant $k_\varphi$ given by:

$$k_\varphi = I\frac{4\pi^2 + \delta^2}{T_d^2} \tag{16}$$

where $\delta$ is the logarithmic decrement and $T_d$ the pseudo-period given by equation (13). In this case, the angular amplitude $\varphi$ is proportional to the voltage amplitude $V$ corrected by the voltage $V_0$ measured when the system is not excited.

The LPM can be used in two measurement modes:
- amplitude resonance mode, which consists in forming a chain of regular short laser beam pulses with a shutter placed in front of the laser head (for instance short shutter aperture of about 0.1 s every second or so), the period being chosen to produce an amplitude resonance (example given in Figure 3a);
- transient mode, which consists in opening the shutter for several seconds to observe the damped oscillation.

In the following, we consider only the transient mode (example given in Figure 3b) where the shutter is opened rapidly (200 ms) and left open for 5 s.

From experimental recordings of damped oscillations as a response to an optical power step, the first two output voltage amplitude maxima $V_{tmax1}$ and $V_{tmax2}$ and the time between them corresponding to the pseudo-period $T_d$ are determined using Matlab® toolbox software ($t$max2 = $t$max1 +$T_d$). The output voltage values must be corrected by the voltage $V_0$ delivered by the PSD for an optical power of 0 mW (here $V_0$ = -0.34 mV). In the following example, $\delta$, $T_d$ and $k_\varphi$ are determined from the measurements of voltage response to a radiation pressure single-step obtained at five different optical powers. The angular spring constant $k_\varphi$ is calculated from equation (16) where $\delta$ is given directly by equation (13). Note that it is not necessary to know the values of optical power $P$ to determine these parameters which are characteristics of the system.
From equation (13), the uncertainty in $\delta$ depends not only on the uncertainty of reproducibility of voltage measurements but also on the voltage uncertainty due mainly to the voltage fluctuations caused by seismic disturbances. The voltage fluctuation is about ± 0.6 mV. For a rectangular probability distribution, the standard uncertainty in the voltage measurement is $u_V$ = 0.35 mV.
The standard uncertainty for $\delta$ determined from the measurements for an optical power between 0.5 and 1 W is 0.026.
The standard uncertainty for $T_d$ is the quadratic sum of the measurement uncertainty for an optical power between 0.5 and 1 W of 0.025 s and the uncertainty of repeatability, reproducibility and hysteresis effect of about 0.01 s, *i.e.* 0.027 s.
From equation (16), the uncertainty for $k_\varphi$ depends on those of $I$, $\delta$ and $T_d$.

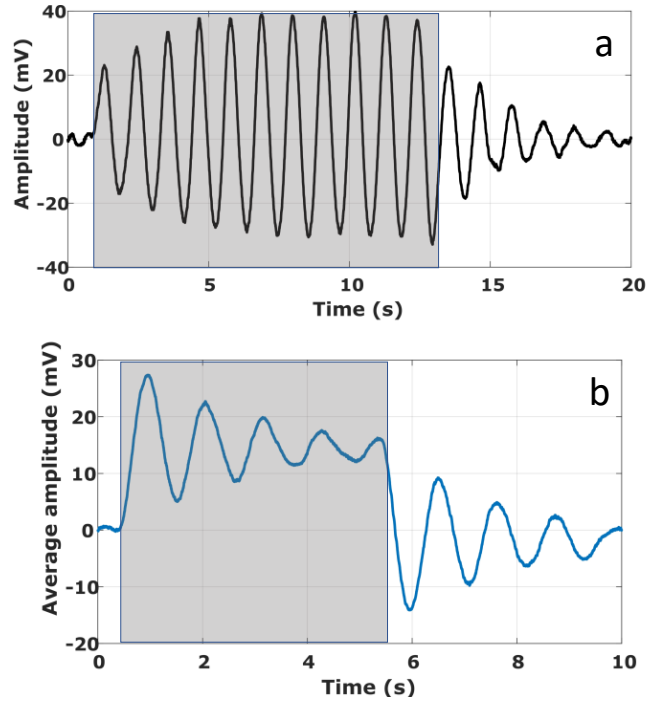

Figure 3: Example of a voltage signal delivered by the PSD sensor obtained:
(a) in amplitude resonance mode (the grey zone corresponds to the time when the laser beam is modulated by the shutter, closed before and after this time);
(b) in transient mode (the grey zone corresponds to time when the shutter is open. It is closed before and after).

Table 1 gives the elements required to evaluate the uncertainty of $k_\varphi$ and shows the main contribution arises from the uncertainties of the moment of inertia $I$ and the pseudo-period $T_d$, all values being expressed in SI units.

|   | Measurand value | Standard uncertainty | Sensitivity coefficient | Standard uncertainty |
|---|---|---|---|---|
| $I$ | $8.32 \times 10^{-8}$ | $2.43 \times 10^{-9}$ | 32.2 | $7.8 \times 10^{-8}$ |
| $\delta$ | 0.182 | 0.026 | $-2.47 \times 10^{-8}$ | $6.4 \times 10^{-10}$ |
| $T_d$ | 1.106 | 0.027 | $-4.84 \times 10^{-6}$ | $1.3 \times 10^{-7}$ |
| $k_\varphi$ | $2.68 \times 10^{-6}$ | - | - | $1.52 \times 10^{-7}$ |

Table 1: Elements to evaluate the uncertainty of the angular spring constant $k_\varphi$. All values in SI units.

### 3.3.2. Sensitivity

For a moment of force of $74 \times 10^{-12}$ N m corresponding to a laser power of 1 W, the theoretical torsion angle at static equilibrium should be 28 µrad. However, the radiation force is reduced slightly due to the imperfect reflection of the mirror ($r < 1$) and the non-normal angle of incidence of the laser beam ($\cos\theta < 1$).

The experimentally determined sensitivity $S$ of the PSD is 1.75 mV/µm. The PSD is placed at a distance $l_d$ 162 mm from the right-angle prism mirror of the detection laser beam. Assuming the torsion angle $\varphi$ is very small and $l_m \ll l_d$, we deduce $\varphi$ from the measured voltage change $\Delta V = V - V_0$ of the PSD output using:

$$\varphi = \frac{\Delta V}{2 l_d S} \qquad (17)$$

At static equilibrium, the moments are balanced according to:

$$\frac{2 r l_m}{c} P_r \left[ \left(1 - \frac{\varphi^2}{2}\right) \cos\theta - \varphi \sin\theta \right] = k_\varphi \varphi \qquad (18)$$

where $P_r$ is the optical power detected by the LPM.

We deduce:

$$P_r \approx \frac{ck_\varphi}{2rl_m\cos\theta}\varphi \approx \frac{ck_\varphi}{4l_d S r l_m \cos\theta}\Delta V = \vartheta \Delta V \tag{19}$$

With a small angle of incidence $\theta$ and over a limited range of optical power where $\varphi$ remains small, $P_r$ is linearly proportional to $\Delta V$.

We determine the uncertainty of the coefficient $\vartheta$ from equation (19) as shown in Table 2 where all values are expressed in SI units. The uncertainty components for $r$ and $\theta$ are negligible compared with the other contributions.

|  | Measurand value | Standard uncertainty | Sensitivity coefficient | Standard uncertainty |
|---|---|---|---|---|
| $k_\varphi$ | $2.68\times10^{-6}$ | $1.52\times10^{-7}$ | $2.43\times10^{+7}$ | 3.7 |
| $l_d$ | 0.162 | 0.005 | -402.1 | 2.0 |
| $S$ | 1750 | 25 | -0.037 | 0.93 |
| $r$ | 0.991 | 0.0005 | -65.7 | 0.033 |
| $l_m$ | 0.011 | 0.00025 | -5921.8 | 1.5 |
| $\theta$ | 0.040 | 0.005 | -2.61 | 0.013 |
| **$\vartheta$** | **65.1** | - | - | **4.6** |

Table 2: Elements used to evaluate the uncertainty of the coefficient $\vartheta$. All values in SI units.

We obtain a voltage sensitivity $\vartheta = 65.1$ (4.6) W V$^{-1}$, *i.e.* a relative standard uncertainty of 7%.

4. Preliminary results and discussion

The laser used for this study is a frequency doubled Nd:YAG laser (Innolight Prometheus 20NE) providing outputs at both the fundamental 1064 nm (~1 W) and the second harmonic 532 nm (~20 mW). The current (from 1 to 2 A) of the laser diode that pumps the NPRO (Non-Planar Ring Oscillator) crystal is used to vary the laser power. In our example, of course the optical power is measured after the photon beam has passed through three diaphragms.

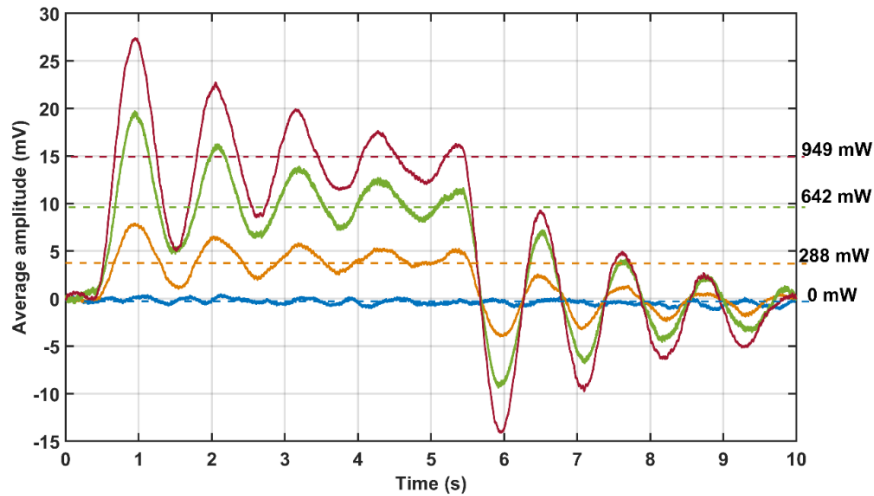

Figure 4: Measurements of voltage responses of the PSD. Each curve corresponds to the average of five signal records for increasing power (up) and five for decreasing power obtained under repeatable conditions for on-off radiation steps of 10 s at three different optical powers. Corrections for offset and drift have been applied. The dashed lines represent the calculated voltage $V_s$.

Figure 4 shows an example of voltage responses to radiation steps in transient mode for different optical powers from five repeated measurements for five increasing powers (up) and five decreasing powers. We observe that the absolute mean value of the voltage response at 1 W is about 15 mV, quite close to the theoretical value.

The maximum amplitude $V_{tmax1}$ of the first oscillation allows one to deduce the stabilization value $V_s$ using:

$$V_s = \frac{V_{tmax1} - V_0}{1 + \exp(-\delta/2)} \qquad (20)$$

From equation (20), the optical power $P_r$ can be determined from equation (19) where $\Delta V = V_s$.

Table 3 gives an example of two series of optical power measurements $P_r$ determined from the measured voltage amplitudes $V_{tmax1}$ for the same seven laser powers $P$ (measured by the commercial optical power meter): one for increasing power, the other for decreasing power. The values of $V_s$ and $P_r$ are calculated using equations (19) and (20) respectively with $\vartheta = 65$ W V$^{-1}$.

| $P$ (mW) | $V_s$ up (mV) | $P_r$ up (mW) | $V_s$ down (mV) | $P_r$ down (mW) |
|---|---|---|---|---|
| 133 | 2.10 | 137 | 2.15 | 139 |
| 288 | 4.29 | 279 | 4.00 | 260 |
| 486 | 7.32 | 476 | 7.38 | 480 |
| 642 | 10.46 | 680 | 948 | 616 |
| 757 | 11.78 | 766 | 11.47 | 745 |
| 856 | 13.10 | 851 | 12.79 | 831 |
| 949 | 14.47 | 941 | 14.58 | 948 |

Table 3: In this example, the voltage amplitudes $V_{tmax1}$ were measured for increasing (up) and decreasing (down) powers. $V_s$ and $P_r$ are calculated from equations (26) and (25) respectively.

5. Potential sources of error and uncertainties in measurement

Several errors limit the accuracy of force measurement using torsion balance techniques [19]. For our torsional magnetic spring system, the main potential sources of systematic error arise from thermal noise, magnetic fluctuations, as well as seismic and gravitational effects. In addition, there is another source of error specific to property of PyC, namely the effect of thermally-excited electrons. On top of these, there are other common metrological sources of error discussed at the end of this section.

5.1. Effect of the thermally-excited electrons

In previous studies [20,21], we used a specific property of pyrolytic carbon to measure laser power namely the effect of magnetic force change when the PyC surface is illuminated by a laser beam.
In the present study, this effect must be eliminated to leave only the effect of the radiation force.
In a first set-up, the right-angle prism mirror was placed directly on the PyC disc.

Figure 5 shows an example of five recordings of the PSD voltage response with the shutter open for 5 s with it closed for 5 s for each optical power at a wavelength of 1064 nm. The aperture time of 5 s in the single-step mode was chosen to limit the influence of any room temperature change and of any internal thermal heating due to IR scattering and reflection by the dielectric mirror.
The general single-step response at the shutter aperture resembles that of either a first-order or an over-damped second-order dynamic system. In addition, we observe rebound effects during the transient phase. The same phenomena are also observed after shutter closure. Moreover, the size of the final voltage (700 mV) is much higher than the theoretical value of about 15 mV.
This disagreement is attributed to thermal diffusion which modifies the magnetic stiffness. Using a thermal imaging camera (Optris OPTPI16-O31T900), we have observed a temperature increase around the mirror of about 5 °C for the first 5 s of irradiation. This means there is thermal transfer via radiation and conduction between the mirror and the PyC disc, which causes the rotational magnetic stiffness to fall sharply. In addition, all the energy reflected by the dielectric mirror is scattered by elements of the box and can heat the PyC disc. To avoid this phenomenon, we must protect the PyC disc against radiation and heat. This is why, to insulate thermally the PyC disc from the mirror and laser beam, the dielectric mirror is placed on a cover glass far away from the PyC disc at about 3 mm using three polyamide balls. Thus, the thermal effect apparent in Figure 5 is not significant in Figure 4.

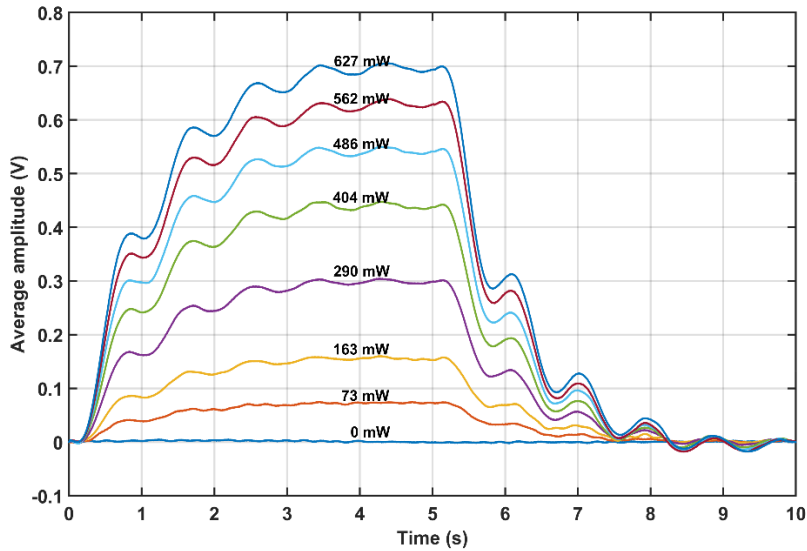

Figure 5: Example of means of five measurements of voltage responses of the PSD device obtained in conditions of repeatability by 5 s irradiation step for different optical powers. Corrections for offset and drift have been applied. Powers are those measured by the optical power sensor.

Even so, the results shown in Figure 5 are interesting. Thus, Figure 6 shows that the response is quasi linear with respect to the optical power shown in Figure 5 when the amplitude of the PSD signal is averaged for 1 seconds between 4 and 5 seconds after the shutter aperture once the signal has become steady enough. At the same time, there is no well-known physical law that can be applied to predict the behavior of the sensor. Consequently, it is not an absolute power meter and, in this case, it is necessary to calibrate the sensor with a calibrated power meter.

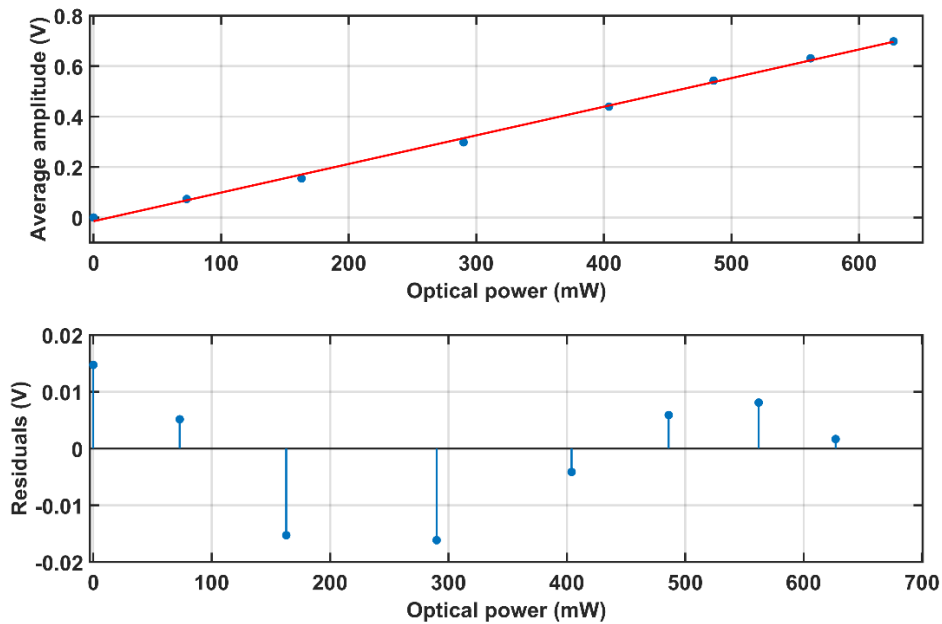

Figure 6: Voltage amplitude delivered by the PSD device averaged between 4 and 5 seconds after aperture as a function of the optical power of the laser beam ($\lambda$ = 1064 nm) measured by the optical power detector. The line is determined by least squares regression from each mean value. The optical power values are those measured through the three diaphragms. The residuals are shown in the lower graph.

## 5.2. Thermal effect

Temperature change also affects magnetic fields. A variation of temperature $\partial T$ causes a variation of magnetic induction $\partial B = B \zeta \partial T$ with a temperature coefficient of remanence $\zeta$ of -0.1 % °C$^{-1}$. In a room where the ambient air temperature was (20.0 ± 1.0) °C throughout the measurements, the relative variation of the induction is about a part in a thousand. In our experiment, the influence of magnetic fluctuations is negligible because the temperature of the magnet array measured by a thermistor is regulated by means of a Peltier module to within ± 4 mK. Thus, we can expect the temperature of the magnet array to be stabilized to better than ± 0.1 °C.

In fact, the largest thermal effect is caused by temperature gradients related to small draughts near the levitated system. These are reduced by housing the device in a small almost airtight box with holes for the beams of the detection laser and the laser under test. Despite this protection, draughts remain one of the main sources of instability of the LPM.

We now consider only molecular bombardment of our torsion device and we assume the device is bombarded isotropically by air molecules. Under these conditions, the limit on the effect of thermal noise can be determined by a Langevin analysis method by applying the principle of equipartition connecting fluctuations and dissipation.

The article by M$^c$ Combie [22] gives an equation for the thermal noise-induced angular fluctuation $\Delta \varphi$ of a static pendulum observed for a duration $\tau$:

$$\Delta \varphi = \pm \frac{\pi}{\omega_0^2} \sqrt{\frac{kT}{I \tau \tau^*}} \tag{21}$$

where $\tau$ is the duration of measurement (observation time) and $\tau^*$ the damping time (time constant) of the pendulum.
From experimental results, $T \approx 300$ K, $\tau^* \approx 1.7$ s and $\tau \approx 5$ s. Under these conditions, we obtain $\Delta \varphi = \pm 7.5$ nrad.

## 5.3. Seismic effects

The second major source of fluctuation is caused by the vertical and horizontal components of vibrations transmitted mechanically to the magnet array. All vibrational modes can couple into the levitated system and lead to increased noise. Speake and Gillies [23], for instance, consider the case of a torsion pendulum with a fibre suspension point driven by a horizontal vibration and obtain the torsional noise using standard Lagrangian methods. They show the fluctuating torsional angle is proportional to the power spectral density and inversely proportional to the natural frequency of the pendulum. In our case, horizontal vibrations (of about 1 Hz) can easily couple into the levitated system for which the horizontal and torsional natural frequencies are about 1 Hz. This effect is greatest when the horizontal component of seismic vibrations lies roughly along the same direction as the incident laser beam.

## 5.4. Gravitational effects

Newton's law states there is an attractive force between two objects, proportional to the product of their masses. If one of them moves with respect to the other (e.g. displacement of the experimentalist with respect to the LPM), the attraction force varies. This fact could modify the LPM behaviour. Gillies and Ritter [18] discuss the case where there is movement of masses in the vicinity of a torsion pendulum leading to a variable torque on it.

In our case, we write an approximate equation for the small torque $\partial \Gamma$ due to a large mass $M$, the centre of gravity of which is located in the same plane $(x,y)$ as the PyC disc.

$$\partial \Gamma = -12 \frac{CGM m_m l_m^2}{r_m^4} \partial r_m \tag{22}$$

where $G$ is the Newtonian gravitational constant, $C$ a coupling coefficient taking the angular dependence of interaction into account ($-1 < C < +1$) and $r_m$ the distance from the mass $M$ to the centre O of the PyC disc. Note that there is no gravitational effect if the centre of gravity of the mass $M$ is located either on the alignment axis of the two mirrors (axis O$y$) or on the symmetry axis of the two mirrors parallel to the $x$-axis (axis O$x$). There is a torque only for the other positions in the $x$-$y$ plane.
For instance, if $C = 1$ for the LPM, we obtain $\partial \Gamma = -15 \times 10^{-15}$ N m for $M = 1$ kg, $r_m = 10$ cm and $\partial r_m = 1$ mm corresponding to a torsion angle change $\Delta \varphi = 0.6$ prad. For a mass $M = 100$ kg at $r_m = 0.5$ m displaced by $\partial r_m = 10$ cm, the torsion angle change $\Delta \varphi = 0.9$ nrad.

Another gravitational effect that might modify the behavior of the levitated system is the global effect on the system mass $m$ which could cause a horizontal displacement of the centre of gravity of the system. For a mass of 100 kg located 1 m away from the levitated system, we obtain a lateral gravitational force $\partial F_g = 1$ pN. Given the approximate value ($43 \times 10^{-3}$ N m$^{-1}$) of the horizontal spring constant $k_x$, the centre of gravity of the levitated system is displaced horizontally by about 24 pm.

In this theoretical example, the gravitational effects on the LPM are negligible.

5.5. Accuracy and uncertainty in measurement

The standard and relative standard uncertainties related to the results presented in Table 3 are shown in Table 4.

| $P_r$ up (mW) | $u_{pr}$ up (mW) | $u_{pr} / P_r$ | $P_r$ down (mW) | $u_{pr}$ down (mW) | $u_{pr} / P_r$ |
|---|---|---|---|---|---|
| 137 | 49 | 37 % | 139 | 11 | 9 % |
| 279 | 55 | 19 % | 260 | 20 | 7 % |
| 476 | 67 | 14 % | 480 | 35 | 7 % |
| 680 | 83 | 13 % | 616 | 45 | 7 % |
| 766 | 90 | 12 % | 745 | 54 | 7 % |
| 851 | 97 | 11 % | 831 | 60 | 7 % |
| 941 | 105 | 11 % | 948 | 69 | 7 % |

Table 4: Standard and relative standard uncertainties of the optical powers given in Table 3 measured using the LPM.

These results were obtained using two methods:
- one based on the law of propagation of uncertainty and the characterization of the output quantity by a Gaussian distribution (evaluation of the standard uncertainty associated with an estimate of the output quantity);
- the other based on the propagation of probability distributions using a mathematical measurement model (Monte Carlo method).

The present example shows, amongst other things, the influence of seismic vibrations. The measurements for increasing power (up) were carried out one morning while there were ground roadworks close to the laboratory. Those for decreasing power (down) were carried out on the afternoon of the same day once the roadworks had stopped. In this case, the relative standard uncertainty is 7%, except for the smallest optical power.

To compare the agreement of the two methods for measuring optical powers $P$ and $P_r$, we calculate the normalized error $E_n$ using formula (23) below assuming the two variables $P$ and $P_r$ to be normally distributed:

$$E_n = \frac{|P - P_r|}{\sqrt{U_P^2 + U_{Pr}^2}} \tag{23}$$

The measured values $P$ and $P_r$ are not significantly different if $E_n < 1$. Given the relative calibration uncertainty of the commercial optical power sensor is about 7 % (manufacturer's data), the calculation for the 14 paired values in Table 3 gives $0.02 \leq E_n \leq 0.4$, except for one paired value $(P, P_r) = (288$ mW , $260$ mW$)$ where $E_n = 0.98$. Apart from this value, all the measured values $P$ and $P_r$ can be declared not significantly different.

These experimental results highlight the influence of seismic vibrations, particularly for the result obtained for 288 mW. For this reason, the series for increasing power was repeated on another day when ambient vibrations were exceptionally low. In this case, the voltage measurement of $V_s$ at $P = 288$ mW was 4.42 mV giving a value of $P_r$ of 287.7 mW, which is very close of the value of $P$ and, consequently, with a very low normalized error of 0.01.

Williamson's method [24] developed with Microsoft Excel software was applied. This method is used for weighted least-squares fitting when both the input and output data have uncertainties that vary from one point to another. In the present case, the measured voltage $V_s$ is the data delivered by the LPM to be calibrated and the optical power $P$ measured by the Thorlabs optical power sensor used as the calibration reference. In addition, it is assumed there is no covariance between the values of $P$. From both experimental measurement series (up and down) corresponding to 14 paired measurements, the software yields a linear relation of the form $P = aV_s + b$. The calculation using Williamson's method gives:

$$P = 65.27 \times V_s + 2.6 \tag{24}$$

where $P$ is expressed in mW and $V_s$ in mV.

In fact, the standard uncertainty $u_b = 6.0$ mW for $b$ is quite large while the experimental covariance cov($a,b$) is found to be negligible. Consequently, it is realistic to assume $V_s = 0$ for $P = 0$. The standard uncertainty for $a$ is $u_a = 0.14$ mW/mV.

Taking an overall uncertainty for $V_s$ of 0.35 mV, the standard uncertainty of the optical power in mW determined from equation (24) is:

$$u(P) = \sqrt{0.02 \times V_s^2 + 522} \qquad (25)$$

For instance, if $V_s = 15$ mV, equation (24) gives $P = 979$ mW with an expanded uncertainty $U(P) = 46$ mW at the 95 % confidence level determined from equation (25), *i.e.* a relative expanded uncertainty of 4.7 %. Note that the expanded uncertainty remains at about 46 mW for all measured values $V_s$ of below 30 mV corresponding to an optical power of about 2 W. The 0.35 mV uncertainty in the voltage measurement is the greatest source of uncertainty in optical power measurement.

Figure 7 presents the graph obtained using Williamson's method with the experimental values (circles), the linear model (solid line) corresponding to equation (24) and the dashed curves corresponding to the standard uncertainty limits. The residuals are also shown.

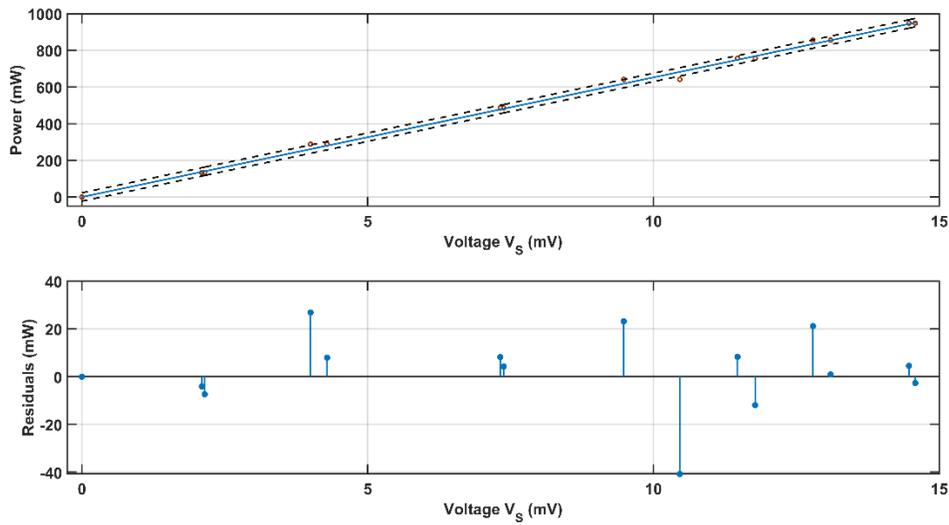

Figure 7: Graph obtained using Williamson's method with the experimental values (circles), the linear model (solid line) corresponding to expression (24) and the curves corresponding to the standard uncertainty limits (dashed curves). Residuals are shown below.

Note that we could extrapolate the linear model up to 10 W since the deflection angle $\varphi$ still remains small at such a power. The value $V_s$ for 10 W would be about 154 mV, corresponding to a deflection angle $\varphi$ of 270 nrad, *i.e.* a spot displacement of the detection laser beam of 88 µm on the surface of the PSD detector. Williamson's method gives a relative expanded uncertainty of 0.6% at 10 W but neglects other components of uncertainty namely that of the reflection angle $\theta$ and the coefficient $r$.

6. Conclusion

We have demonstrated a new optical power meter (laser power meter or LPM) based on the measurement of radiation pressure with a diamagnetic spring. The results obtained in transient measurement mode with a very simple and inexpensive experimental set-up lead to the following findings:
- The linear model validated in the range from 0.1 to 1 W gives a relative expanded uncertainty for optical power measurement from about 20% to 5% in the range 0.3 to 1 W.
- The main uncertainty, related to voltage measurement, yields a minimum uncertainty of 46 mW at the 95 % confidence level.
- The measured values $P$ (integrating sphere photodiode power sensor) and $P_r$ (LPM) are not significantly different, the relative deviation between $P$ and $P_r$ being 10% or less.
- The extrapolation of the linear model up to 10 W looks a promising approach for measuring optical powers with an accuracy of a few percent.

The LPM is sensitive to vibrations, draughts and temperature changes resulting in deviation if the optical power is measured at different times.

The device demonstrated is an *absolute* optical power meter which means one needs only to determine some of its characteristics independent of the optical power. This characterization consists in the measurement of the total moment of inertia $I$, the distance between the right-angle prism mirror and the PSD sensor $l_d$, the distance between a prism mirror and the rotation axis $l_m$, the coefficient of reflectivity and absorption $r$, the pseudo-period $T_d$, the logarithmic decrement $\delta$, the sensitivity of the PSD sensor $S$ and the angle of incidence $\theta$.

Moreover, both the accuracy and sensitivity of the LPM could be improved.

To improve the accuracy, one needs to reduce the uncertainties of the moment of inertia $I$ of the levitated system and of the measurements of the voltage and the pseudo-period $T_d$. To reduce the voltage measurement uncertainty, the LPM should be protected against vibrations. In addition, a suitable enclosure would protect it efficiently against draughts. The PyC disc and the magnet array must be maintained at constant temperature, in particular, no infrared radiation should fall on the disc while the temperature of the magnet array must be servo-controlled e.g. using a Peltier module.

The LPM sensitivity could be increased by reducing the angular spring constant $k_\varphi$, e.g. by reducing the size of the holes in the PyC disc. Note that $k_\varphi$ could be also reduced by placing the LPM under vacuum to reduce the damping constant $\gamma_\varphi$. In this case, however, the three damping constants $\gamma_x$, $\gamma_y$ and $\gamma_z$ would also be reduced, making the LPM far more sensitive to vibrations.

Several authors [25,26] use an angular amplification method based on multiple reflections. We could use it to amplify the angle of deviation of the laser beam whose optical power must be measured. However, while the radiation force of the laser source would be significantly increased, the angular response would no longer be linear. This is why for an amplification using several reflections, we would need to develop a feedback control loop, for instance, one based on an electrostatic actuator. This would allow us to improve the linearity of response and the signal-to-noise ratio and would be a good way for measuring weak optical powers i.e. below 0.5 W.

In future work, we shall study the behavior of the LPM in amplitude resonance mode which consists in modulating the incident laser beam at low frequency (around 17 Hz).

There are also other possible applications of PyC for measuring weak contact forces [27,28]. Indeed, the present device could be used for this but only using a linear component of magnetic spring along either a single horizontal axis or vertical axis.

Acknowledgements

The authors thank Dr Mark Plimmer for his kind reading of the manuscript and Dr Jean-Pierre Wallerand for helpful discussions.